# Sustainable Border Control Policy in the COVID-19 Pandemic: A Math Modeling Study

Authors: Zhen Zhu[1], Enzo Weber[2], Till Strohsal[3], Duaa Serhan[4] *

## ABSTRACT

**Background:** Imported COVID-19 cases, if unchecked, can jeopardize the effort of domestic containment. We aim to find out what sustainable border control options for different entities (e.g., countries, states) exist during the reopening phases, given their own choice of domestic control measures and new technologies such as contact tracing.

**Methods:** We propose a SUIHR model, which represents an extension to the discrete time SIR models. The model focuses on studying the spreading of virus predominantly by asymptomatic and pre-symptomatic patients. Imported risk and (1-tier) contact tracing are both built into the model. Under plausible parameter assumptions, we seek sustainable border control policies, in combination with sufficient internal measures, which allow entities to confine the virus without the need to revert back to more restrictive life styles or to rely on herd immunity.

**Results:** When the basic reproduction number $R_0$ of COVID-19 exceeds 2.5, even 100% effective contact tracing alone is not enough to contain the spreading. For an entity that has completely eliminated the virus domestically, and resumes "normal", very strict pre-departure screening and test and isolation upon arrival combined with effective contact tracing can only delay another outbreak by 6 months. However, if the total net imported cases are non-increasing, and the entity employs a confining domestic control policy, then the total new cases can be contained even without border control.

**Conclusion:** Extremely strict boarder control measures in China, which has eliminated domestic spreading, are justifiable. However such harsh mesasure are not necessary for other places. Entities successfully confining the virus by internal measures can open up to similar entities without additional border controls so long as the imported risk stays non-increasing. Opening the borders to entities lacking sufficient internal control of the virus should be exercised in combination with pre-departure screening and tests upon arrival.

## BACKGROUND

To "bend the curve" of the first wave of COVID-19, authorities closed their borders, put stay-at-home orders in place, and enforced social distancing. Afterwards, governments in many countries

[1] Zhen Zhu, Ph.D., oneworld Management Company
[2] Prof. Dr. Enzo Weber, Institute for Employment Research Nuernberg and University of Regensburg
[3] *Corresponding author:* PD Dr. Till Strohsal, German Federal Chancellery and Freie Universität Berlin.
[4] Duaa Serhan, Ph.D., American Airlines

* The opinions in this paper are those of the authors and do not necessarily reflect the views of **one**world Management Company, the German Federal Chancellery, or American Airlines.

and territories have started the phase of reopening to recover their economy. Thereby, when and how to reopen the borders has become a heated topic. While there is little doubt that travel and tourism contribute greatly to the economy, reopening the border in an unstructured manner is likely to increase the risk of a spike of cases and new lockdown orders.

While stricter control measures are more effective in reducing the imported cases, they are also more costly to administer and less favorable for the general public. For example, China has implemented an extremely strict border entrance policy. Travelers must first check-in 14 continuous days on a mobile app to obtain a green health code, and provide a recent nucleic acid negative test certificate and an IgM anti-body negative test certificate pre-departure to the Chinese embassy in the origin country[1,2]. Upon arrival, travelers must undergo immediate tests, followed by a 14-day institutional mandatory quarantine. At the end of the quarantine period, only those who provide another negative test are released. On top of that, the number of international travelers into China is also constrained[3]. While these harsh measures worked in China, they are not necessarily optimal for other places (if even possible). Therefore, when assessing the risks associated with opening the borders against other social and economic interest, it is crucial for governments to understand what the least restrictive but sustainable options are, which are still consistent with their own domestic policy targets.

This paper proposes a holistic approach to study border control policies against COVID-19 during reopening phases. Border control measures are studied in conjunction with domestic control measures, with a special focus on contact tracing. Domestic control strategies against corona vary strongly across countries. We classify government entities (e.g., states, or countries) in the world into three major groups based on their internal strategies; G1 entities able to completely eliminate domestic cases and then started to reopen, e.g., China and New Zealand. G2 entities aim to keep COVID-19 under check until effective vaccines are widely deployed and have started to reopen before the virus was completely eliminated. Many European Union countries belong to this group, where the strictness of confinement measures varies according to the current infection dynamics. G3 entities are those going towards herd immunity either strategically or involuntarily. Our results show that the necessary strictness for border control policies to be sustainable varies strongly for the three types of entity groups.

Optimal policies facing the trade-off between health and economic damage has always been a heated topic[4]. It has been confirmed that social distancing and lockdowns were successful in bringing down the number of new infections[5-9], but also lead to labor market slump[10]. Similarly, a combination of targeted policies with contact tracing is shown to be very effective in preventing a high death toll[4,11]. Tracing together with local lockdowns is shown to be successful in China and Vietnam[12], but may be difficult to exercise elsewhere. A more acceptable trace and isolate method has been developed with the help of high-tech mobile apps. Here, it is crucial that digital contact tracing is fast enough to break infection chains[13], and high proportions of contacts are successfully traced[14-17]. The international spreading of the virus has been studied in the outbreak phase[18]. Travel restrictions are shown to be necessary and effective in interrupting virus outbreak[19,20]. Border control measures, such as syndromic cross-border screening, are found to modestly delay an outbreak from imported cases[21]. Dickens et el. studied the effectiveness of test-isolate and quarantine at the border, and showed that even the 14-day mandatory quarantine alone is not enough to completely stop imported cases[22].

The existing research has produced valuable insights on border control policies in terms of reducing the imported cases. However, it remains to be investigated whether any border policy can actually guarantee that the epidemic remains under control in the destination entities. Our paper contributes to filling this gap. We discuss the necessary conditions on border controls to avoid future lockdowns given the domestic control policies, and limits on incoming traffic from different risk sources tailored to meet different government objectives. Using results from the literature to calibrate our model, we provide governments guidance on sustainable border control policies.

# METHODS

This paper proposes an extension to the Markov transition based SIR and SEIR models that have long been used to model epidemic spreading[23] and control measures such as lockdowns and vaccines[24]. Our approach can be described as a **S**(usceptible) **U**(nidentified infectious) **I**(dentified infectious) **H**(ospitalized) **R**(ecovered) model. In this study, we use the SUIHR model to evaluate combinations of border and domestic control measures against COVID-19, during a reopening phase and before a vaccine is widely deployed. The SUIHR model framework can include multiple entities, reflecting different government bodies with different containment measures. The model framework is tailored to study the sustainability of border control strategies combined with different levels of internal controls and 1-tier contact tracing and isolation. A strategy is considered *sustainable* when it allows an entity to control the virus over time without further tightening the initial internal confinement measures. Additionally, levels of border openness toward different incoming sources can be optimized conditionally on new-case targets and medical resource capacity using our framework.

## THE SUIHR MODEL

The SUIHR model has the following features. First, we explicitly model the importation of infected individuals from one entity to another so that the effect of border control policies can be studied. Second, the model focuses on the undetected spreading of virus from mainly pre-symptomatic or asymptomatic people. We deliberately split the Infectious state in the SIR models into **U**nidentified Infectious and **I**dentified Infectious states, to account for the characteristics of a reopening phase compared to the initial outbreak control phase. For example, the danger of COVID-19 is now well known, and people have the awareness to avoid spreading once they are identified as infected. Testing capacity increased, most people who show symptoms can get a diagnose quickly and will be isolated. Latent/Exposed states that exist in most SEIR models are considered as part of our UI state because pre-symptomatic spreading is known to occur with corona virus[25]. Third, contact tracing is built into the model. During a reopening phase, "trace & isolate" should be preferred over lockdowns as it is less disturbing to daily life and hence more sustainable. Fourth, constraints such as new case targets and medical resources can be incorporated into the model framework as part of the policy optimization process. Solutions to the model can be obtained via linear programming.

In general, unidentified and identified infectious, and recovered individuals can either be free or isolated/quarantined. However, with our focus on reopening phases, we optimistically assume that anyone who has been identified as infectious will be isolated immediately to avoid virus spreading. Hospitalized people are also automatically isolated in the model. Therefore, only the Unidentified Infectious cohort is divided into two mobility states: **F**ree and **Q**uarantined. In contrast to the initial outbreak control phase, we do not consider block quarantine on susceptible people. Hence, the marginal positive effect on reducing the virus spreading by isolating susceptible people is negligible. Instead, the focus is on quarantining the unidentified infectious individuals.

The proposed SUIHR model is in discrete time, each time epoch is defined to be one week. Weekly averages are commonly used to measure COVID-19 statistics and also avoid difficulties connected to day-of-the-week effects. Government actions are taken quite quickly in times of COVID, but a weekly frequency is sufficient in order to guide policies we are occupied with. From a model perspective, the weekly framework allows the integration of constraints during optimization which opens up numerous additional analytical possibilities. A different transitional probability distributional assumption from the SIR/SEIR model is made to incorporate the states with non-exponential stay time[26]. Hence, the Identified Infectious and Hospitalized states are further divided into Stage 1 and Stage 2. For more details on the model states, see the Appendix.

## CONTROL POLICY PARAMETERS

The expected number of people an infectious person can spread the virus to in a time epoch, denoted by $r$, is used to measure the effectiveness of specific control policies. For example, allowing gathering of large groups of people would increase $r$, while enforcing mask wear would reduce it. Assuming only people in the state of unidentified infectious and free to move can travel, only a small percentage of travelers are infectious upon arrival. This percentage, denoted by $\alpha$, is mainly determined by the COVID-19 severity at the origin and the pre-travel screening policies (e.g., people must provide RNA/antibody test results prior to travel). Also, people could potentially get infected during the trip, hence $\alpha$ upon arrival at the destination entity could be higher than at departure. When a test upon arrival is administered, $\tau$ (percentage) of the infected travelers will be identified and isolated. We use $\tau = 0.91$ with mandatory test for everyone upon arrival[22]. In the absence of such test, $\tau = 0$. When there is a quarantine policy upon arrival, a certain percentage of the unidentified infected travelers, denoted by $\beta$, will go through a strict quarantine and won't spread the virus. The strictest 14-day mandatory institutional quarantine will have $\beta = 1$, while self-administered home quarantine is assumed to be have $\beta = 0.6$ as there is no guarantee everyone will strictly abide by the rules. Contact tracing is also explicitly modeled. When someone is identified as infected, contact tracing will reveal a percentage of the people (denoted by $\theta$) the newly identified person has directly contacted. Those contacts will then be quarantined immediately. In the context of this paper, by contact tracing we will refer only to the 1-tier tracing policy (i.e., only direct contacts of an identified case can be traced and isolated).

In order to contain the virus spreading, the effective reproduction rate must be less than 1. Herd immunity is one way to achieve this. However, herd immunity is unlikely to be the solution for COVID-19[27]. Let $\hat{\mathcal{R}} = P_U^U + (1 - P_U^{I_1} \cdot \theta)$ denote the hypothetical reproduction number of the virus in an entity with contact tracing, assuming everyone is susceptible ($P_U^U$ and $P_U^{I_1}$ are transitional probabilities, see Table 1 for their values). $\hat{\mathcal{R}}$ is a measure of an entity's internal containment policy, and is to be distinguished from the observed reproduction number. For example, if the whole population is immune, the observed reproduction number will be 0 regardless of $\hat{\mathcal{R}}$. When $\hat{\mathcal{R}} < 1$, an entity has a confining policy for COVID-19, without the need to rely on herd immunity or vaccine. G1 entities are opening-up internally and pushing domestic life back to the pre-pandemic "normal". Hence, for G1 entities, $\hat{\mathcal{R}} > 1$ (even though the observed reproduction number is 0). G2 entities aim to keep COVID-19 under check (i.e., $\hat{\mathcal{R}} < 1$) until effective vaccines are deployed. G3 entities have $\hat{\mathcal{R}} > 1$ with growing new cases. We group entities based on $\hat{\mathcal{R}}$ to study the sustainability of policies. Nevertheless, the allowance of incoming traffic from different entities are optimized taking their actual incidences into account via the parameter $\alpha$.

## DATA

The data used to estimate the transitional probability matrix for the SUIHR model came from existing studies on COVID-19[28-34]. We acknowledge that the literature regarding COVID-19 is constantly developing and hence the parameters used in this paper are subject to change as better knowledge regarding the virus is available. However, the qualitative interpretation of the results in this paper would be applicable even when the parameters would differ. Table 1 shows the estimation of the transitional probability matrix, and the Appendix provides more details. Without loss of generality, we choose the main and second entity of interest in simulation to have a population of 80 and 50 million people, respectively. We assume 2% of the unidentified infectious people would travel at each time epoch. To put this number into perspective, according to IATA traffic data, in 2019, the number of average weekly international travelers was little less than 2% of the total population in Germany. We assume unidentified infectious people have the same probability of traveling as the general population. This would provide about 10 to 20 daily imported cases to each entity initially.

## RESULTS

Our study focuses on border control policies that can keep $\hat{\mathcal{R}} < 1$ while providing most flexibility in life. We find that when the basic reproduction rate of an infection exceeds a certain threshold, even perfect contact tracing (i.e., $\theta = 1$) alone is not enough to contain the spreading (see Appendix Lemma 1). For COVID-19, we estimate that when $R_0 > 2.5$ (i.e., $r > 1.875$), $\hat{\mathcal{R}} > 1$ even with perfect contact tracing. When the virus first broke out, the basic reproduction rate has an average estimation of $R_0 \approx 3.28$ without any interventions[35]. The rationale behind this is that COVID-19 is possible to spread without the carriers showing symptoms, hence (1-tier) contact tracing is not 100% effective against spreading by asymptomatic and pre-symptomatic cases. Although contract tracing alone cannot bring life back to normal, it is possible to alleviate other measures. Assuming a base reproduction ($R_0$) rate of 2.8 in pre-pandemic normal life with no interventions, the theoretically highest gain from contact tracing would be to achieve a realized reproduction rate of 1.27. As shown in Appendix Figure A.6 and A.7, with good tracing ($\theta = 0.8$), we can allow life going back to $r = 1.4$ whereas without it, the situation will be out of control very quickly when $r > 0.8$. Such gains might be very helpful in relaxing alternative containment measures.

This means that G1 entities will not be able to confine the imported cases from spreading with contact tracing alone. In our simulation study, we assume a good trace & isolation implementation ($\theta = 0.8$), together with a strict pre-departure screening with only 10 cases per week entering the border. For border control measures, we consider the second-best option with mandatory test upon arrival with isolation on positive test cases, and a 14-day home quarantine for everyone else ($\tau = 0$, $\beta = 0.6$). Under these measures, only less than 1 case per week can sneak into the entity undetected. Such tightness of control would be ideal in reality. We assume another lockdown would happen when identified new cases top 40 thousand a week (ca. 6000 per day). Figure 1 shows that even these tight border controls are not enough to prevent another outbreak but only delay it for about 26 weeks (23 weeks without the home quarantine), another lockdown would have to happen if tighter measures are not introduced. This is consistent with the existing studies showing that reducing traveling rates between interconnected cities can delay but not prevent the arrival of an outbreak from an infected source[21,36]. The extreme tight quarantine policies China is implementing for international travelers can be seen in the light of these results.

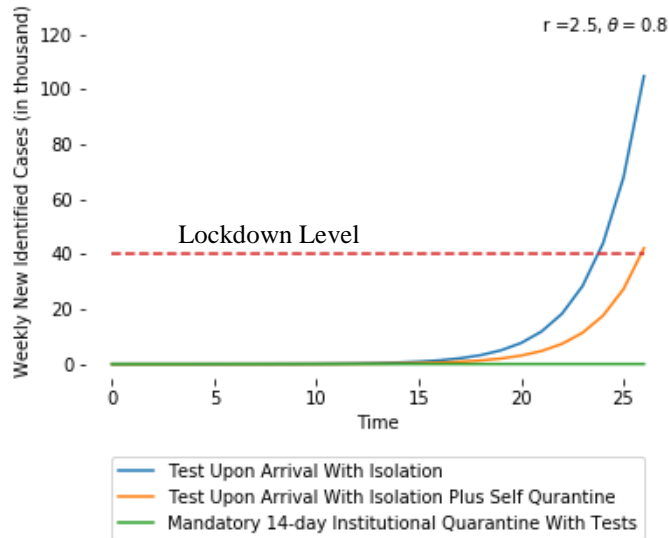

Figure 1: Impact of Mandatory Quarantine for G1 Entities

G2 countries, given a sufficient *confining domestic control policy* (i.e., $\hat{\mathcal{R}} < 1$ without the imported cases), can keep the curve flattened, so long as the imported risk is not growing over time (see Appendix Lemma 2). Assume that the infectious free people have the same probability of traveling as the healthy people, it is possible for G2 entities to open borders towards G1 and other G2 entities. In these two cases, screening and quarantine are not required to keep the virus spreading in check. Nevertheless, those measures can significantly reduce the number of new cases. We simulate the scenario that two G2 entities have complete open border towards each other and no quarantine or screening policies are in place for travelers between them. Figure 2 shows that even with open border, both entities are experiencing decreasing cases very similar to close border policies between each other. We further simulate infections during the trip by doubling the number of infected people upon arrival comparing to the numbers at departure. The simulation suggests that even with a high infection rate on the road, both entities can withstand the imported risk and maintain a decreasing trend for both the identified new cases and the number of hospitalized patients over time.

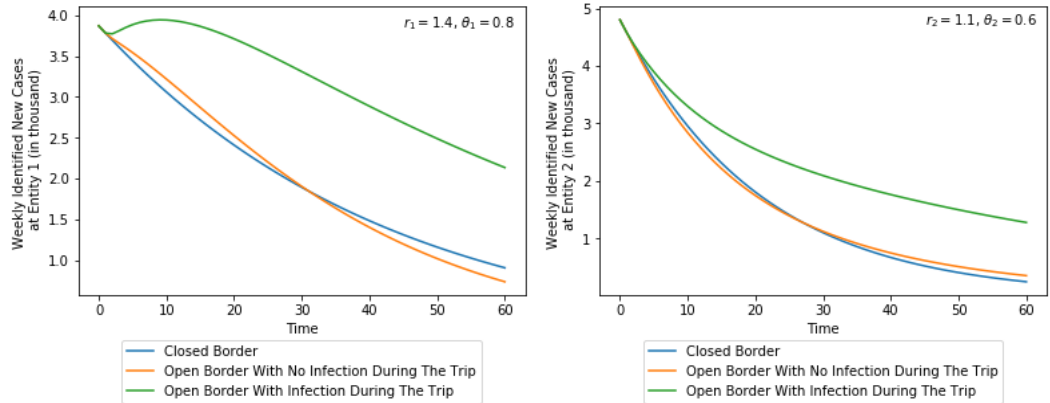

Figure 2: Interaction between two G2 Entities

The same conclusion cannot be drawn for a G2 entity for opening its border to a G3 entity. G2 entities' effort to contain the virus outbreak can be jeopardized by the increasing imported cases from G3 (Appendix Figure A.10). Additional border control policies must be applied toward travelers from G3 entities. Indeed, the effectiveness of travel bans applied to countries with high disease incidence has been demonstrated[37]. In case that G2 (and G1) entities decide to open up to G3 entities, and mandatory quarantine with tests are not a viable or preferable option, our study shows that combining pre-departure screening and test upon arrival may be the answer. In the simulation study, we group all the high-risk entities with increasing cases into a single G3 entity from the view of a G2 entity. We examine the effectiveness of the 72 hours pre-departure screening containing either a single nucleic acid test or antigen test, which could have up to 20% false negative according to the FDA requirements [38,39] (i.e., reduce imported cases by 80%). We use the maximum false negative rate as a conservative assumption on the test sensitivity, to ensure the results are not overstated. Test upon arrival with isolation ($\tau = 0.9$); and self-quarantine for 14-day ($\beta = 0.6$) are also studied along with the combination of the three control measures. Figure 3 shows individual measures when implemented alone may not be able to avoid another lock down within half a year when vaccines are still not widely deployed. However, the combination of pre-departure screening with test-and-isolate and/or self-quarantine could manage the risk under the target.

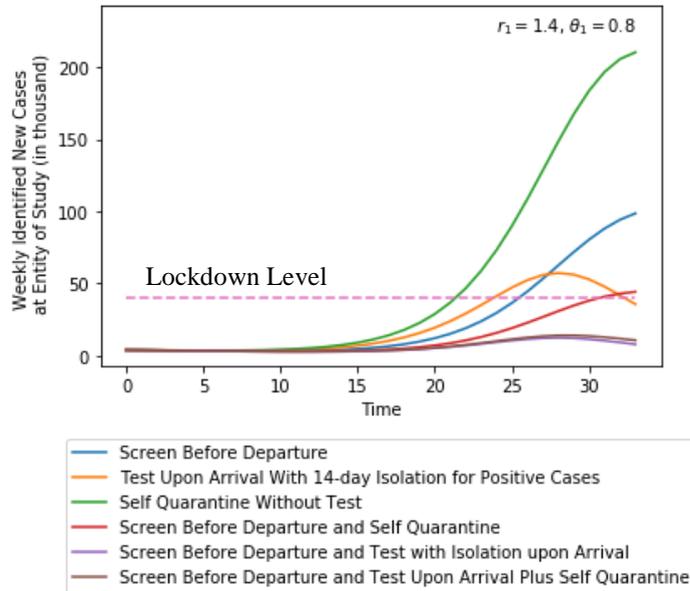

Figure 3: Effectiveness of Different Control Policies toward G3 Travelers

When opening up to multiple entities consisting both G2 and G3, it is crucial to keep the total number of imported cases under a constant level. Figure A.8 in Appendix shows that even without any border control, at a constant 1000 imported case per week, a G2 entity can keep the infection under control (albeit at a relatively high level of 14000 new cases per week or 2000 new cases per day).

## CONCLUSION

A new model is proposed in this paper to study the sustainability of border control policies together with other domestic measures and contact tracing against COVID-19, during a reopening phase in an on-going pandemic. Entities in the world are grouped into three groups based on their states of the domestic epidemic control (G1: eliminating the virus, G2: controlling the virus, G3: witnessing continued spread of the virus). Using our model, we can draw several policy-relevant conclusions and recommendations. First, the effectiveness of border closures also depends on the domestic containment measures in combination. Contact tracing with isolation is an effective method to reduce the reproduction rate, but still some further domestic restrictions are needed. Second, G2 entities can open up to other G2 entities without additional border controls as long as the imported risk stays non-increasing. The level of openness can be determined via linear programming using our model, taking the entities' new case targets or medical resources as constraints; see Appendix for technical details of the LP formulation and solutions. Once the initial level of openness is implemented, we recommend that G2 entities closely monitor the imported cases. While that would be possible by testing all travellers upon arrival, such restrictive and costly measures might be undesired. An alternative would be to make use of random sampling or data from individual tests to find out whether positive cases are travel related. When imported cases are approaching the threshold, border controls should be tightened. Third, when opening the borders to G3 entities, pre-departure screening in combination with test-and-isolate upon arrival can be used in place of the more extreme quarantine interventions. Fourth, G1 entities should impose mandatory institutional quarantine to travelers from any entity that has not eliminated domestic spreading of the virus. Pre-departure screening is also needed when spaces for institutional quarantine are limited. The extreme measures China is implementing is sensible against this background.

| Transitional Probability | Assumed Value | Description and References |
|---|---|---|
| Probability of transition from unidentified infectious to identified infectious, Stage 1 ($P_U^{I_1}$) | 60.0% | This set of parameter means that 80% of the infected people will eventually show symptom with 60% of them showing the symptom in the first week (Lauer, Grantz [29]). Also 20% of the infected will be asymptomatic throughout the course of infection, this is in line with the findings in Streeck, Schulte [31]. |
| Probability of transition from unidentified infectious to recovered ($P_U^R$) | 15.0% | |
| Probability of remaining as unidentified infectious another week ($P_U^U$) | 25.0% | |
| Probability of transition from identified infectious Stage 1 to Stage 2 ($P_{I_1}^{I_2}$) | 71.8% | According to Lechien, Chiesa [30], the mean duration of COVID-19 symptoms of mild-to-moderate cured patients was 11.5 ± 5.7 days. Assuming normal distribution, we estimate 21% of the symptomatic cases can recover with in 7 days. Among patients who developed severe disease, the medium time to dyspnoea from the onset of illness or symptoms ranged from 5 to 8 days[28]. Zhou, Yu [33] also reported the median time from illness onset to dyspnoea is 7 days with the first quarter quantile to be 4 days and third quantile to be 9 days (i.e., IQR being 4 to 9 days). Furthermore, 19% of the total symptomatic cases are severe to critical and would require hospitalization[34]. |
| Probability of transition from identified infectious Stage 1 to hospitalized Stage 1 ($P_{I_1}^{H_1}$) | 9.5% | |
| Probability of transition from identified infectious Stage 1 to recovered ($P_{I_1}^R$) | 18.7% | |
| Probability of remaining as unidentified infectious for 3 or more weeks ($P_{I_2}^{I_2}$) | 27.0% | |
| Probability of transition from identified infectious Stage 2 to hospitalized Stage 1 ($P_{I_2}^{H_1}$) | 9.6% | |
| Probability of transition from identified infectious Stage 2 to recovered ($P_{I_2}^R$) | 63.4% | |
| Probability of transition from hospitalized Stage 1 to 2 ($P_{H_1}^{H_2}$) | 70.0% | The median hospital length of stay is 11 days with IQR being 7 to 14 days [33]. The overall fatality rate here is 2.5% for symptomatic cases, which is well within the range of various studies[32]. |
| Probability of transition from hospitalized Stage 1 to recovered ($P_{H_1}^R$) | 25.0% | |
| Probability of transition from hospitalized Stage 1 to death ($P_{H_1}^D$) | 5.0% | |
| Probability of remaining in hospital for three or more weeks $P_{H_2}^{H_2}$ | 13.0% | |
| Probability of transition from hospitalized Stage 2 to recovered ($P_{H_2}^R$) | 77.0% | |
| Probability of transition from hospitalized Stage 1 to death ($P_{H_2}^D$) | 10.0% | |

Table 1: Transitional Probability Matrix Estimation